\documentclass[reprint,pra,showpacs]{revtex4-1}
\usepackage[usenames]{color}
\usepackage{bm}
\usepackage{amsmath}
\usepackage{amssymb}
\usepackage{epsfig}
\usepackage{graphics}

\newcommand{\bee}{\begin{eqnarray}}
\newcommand{\ene}{\end{eqnarray}}
\newcommand{\lan}{\langle}
\newcommand{\ran}{\rangle}
\newcommand{\eqb}{\begin{equation}}
\newcommand{\eqe}{\end{equation}}
\newcommand{\rd}{\mathrm{d}}

\newcommand{\br}{{\bf r}}

\newcommand{\vare}{\varepsilon }

\begin{document}

\title{Control of the Bose-Einstein condensate by dissipation. Nonlinear Zeno effect}

\author{ V. S. Shchesnovich$^1$ and V. V. Konotop$^{2}$ }

\affiliation{ ${}^1$Centro de Ci\^encias Naturais e Humanas, Universidade
Federal do ABC, Santo Andr\'e,  SP, 09210-170 Brazil,\\
$^2$Centro de F\'isica Te\'orica e Computacional, Universidade de Lisboa,   Avenida Professor Gama Pinto 2, Lisboa 1649-003, Portugal;
Departamento de F\'isica, Faculdade de Ci\^encias, Universidade de Lisboa, Campo
Grande, Ed. C8, Piso 6, Lisboa 1749-016, Portugal}

\begin{abstract}

We show that controlled dissipation  can be used as a tool for exploring
fundamental phenomena and managing mesoscopic systems of cold atoms and
Bose-Einstein condensates.  Even the simplest boson-Josephson junction, that is, a
Bose-Einstein condensate in a double-well trap, subjected to removal of  atoms from
one of the two potential minima allows one to observe such  phenomena as the
suppression of losses and  the nonlinear Zeno effect. In such a system the
controlled  dissipation can be used to create  desired macroscopic states and
implement controlled switching among different quantum regimes.
\end{abstract}
 \maketitle

\section{Introduction}

Universal properties of  condensed atomic gases as  systems hosting spectacular
nonlinear phenomena like instabilities and collapse, solitons and shock waves,
localized modes, vortices and self-trapping and delocalizing transitions, nowadays
are well known~\cite{book}. These   phenomena constitute the properties of
Hamiltonian systems, where dissipation is considered as an undesirable destructing
factor. This role of  dissipation can be inverted, if a system possesses an
intrinsic mechanism balancing the losses and giving origin to stable dissipative
structures \cite{diss1}. Alternatively, the dissipation can lead to a constructive
effect when its action is limited in time, allowing one to generate diverse
nonlinear excitations~\cite{dis_def}, or when it has a nonlinear
origin~\cite{Victor} supporting localized patterns. These effects appear on the
macroscopic scale. Study of the dissipative decay   can reveal also the microscopic
quantum properties of the atomic gases~\cite{Burt} and even inhibit the losses of
atoms, by inducing strong correlations due to a large imaginary scattering
length~\cite{diss_loss}.

\begin{figure}[htb]
\begin{center}
\vskip 1.75cm
\epsfig{file=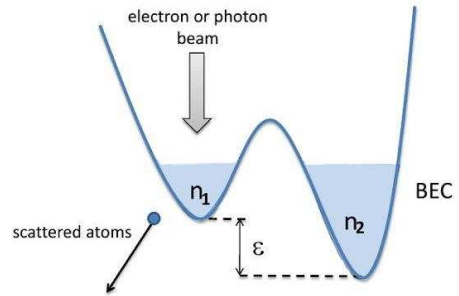,width=0.35\textwidth} \caption{(Color online) A schematic
representation of the setup.  Atoms  are removed from the left potential well
through scattering process by an electron or laser beam.}
\label{pot}
\end{center}
\end{figure}

The purpose of this article is to show that  there are various dissipative regimes
in systems of cold atoms and Bose-Einstein condensates (BECs) loaded in a multiwell
trap and that removal of atoms can  serve as a tool for exploring  the fundamental
quantum phenomena. In particular, we find that the  dissipation can inhibit losses
and allows one to manage the so-called Macroscopic Quantum Self-Trapping state
(below simply the self-trapping state) \cite{MFDW,MFDW2}. Moreover we observe  the
inhibition of losses due to atomic interactions which can be termed as the
macroscopic nonlinear Zeno effect.

Specifically, we consider the case of a BEC in a double-well potential whose
Hamiltonian dynamics is well understood theoretically and has been a subject of
fundamental experiments and numerous potential applications. The basic model, the
well-known   boson-Josephson junction~\cite{Legg,GO},  is also analogous to a
nonrigid quantum pendulum. In particular, it was already used for observation of
the macroscopic quantum tunneling and  self-trapping~\cite{TunTrap}, whereas the
future proposals include also the atomic Mach-Zehnder interferometer~\cite{AtInt},
the sensitive weak force detector~\cite{JM} and the atomic interferometer on a
chip~\cite{AtSqz}. It was previously used also to study  thermal {\it vs} quantum
decoherence~\cite{GK}.

In our setup, schematically depicted in Fig.~\ref{pot},  we introduce an additional
element -- the controlled removal of atoms from one of the two wells of the
potential -- and consider the effect of the dissipation, showing that the actual
rate of atom losses has a dramatic dependence on the dissipation coefficient and on
the atomic interactions strength.

We notice that in the earlier experiments~\cite{FGR}, the Zeno effect was observed
in a system of cold atoms loaded in an accelerating optical lattice with the
magnitude of the acceleration varying in time. The macroscopic manifestation of
Zeno and anti-Zeno effects in the Josephson junctions subjected to time-dependent
perturbations was also studied \cite{BaKuKo}. Also, our statement of the problem,
can be further developed to describe the experimental setup used in \cite{CPZeno}
for  the observation of pulsed and continuous Zeno effects in an externally driven
mixture of two hyperfine states of a ${}^{87}$Rb BEC, using the destructive
measurement of the population of the states.  However, this happens in a binary
mixture with nonlinear interactions between the components, while separation of the
two subsystems by a potential barrier in our setting induces  only a linear
coupling between them (i.e. the populations in the two wells) due to the quantum
tunneling.

The organization of the article is as follows. In Sec.~\ref{sec:master}  we deduce
the master equation describing our system. An exact solution is found in
Sec.~\ref{sec:noninteract} for noninteracting atoms. The nonlinear Zeno effect is
described in Sec.~\ref{sec:nonlin_Zeno}. Sec.~\ref{sec:switch} is devoted to the
description of several switching regimes which are induced and controlled by the
dissipation. In the Conclusion (Sec. VI) we summarize our results and present an
outlook.

\section{The master equation}
\label{sec:master}

We use the  simplest  boson-Josephson Hamiltonian \cite{QHAM,AB,GO} describing BEC
in   an asymmetric double-well potential $V(\br)$, that is,
\eqb
H = -J(a_1^\dag a_2 + a_2^\dag a_1) + \vare{n_1} + U_1n_1^2 +U_2n_2^2,
\label{EQ1}
\eqe
where $a_j$ and $a_j^\dag$ ($j=1,2$) are boson operators for the wave functions
$\varphi_{1,2}$ localized at the potential minima, $n_j = a^\dag_ja_j$, $J$ is the
single atom tunneling rate, $\vare$ is the zero-point energy bias,  $U_{1,2}=
g/2\int \rd^3\br \varphi_{1,2}^4$, where $g=4\pi\hbar^2 a_s/m$, $a_s$ is the s-wave
scattering length, and $m$ is the atomic mass (for a  trap with the two
quasidegenerate lowest energy levels, $U_1\approx U_2=U$ with a good accuracy).

The  controlled removal of atoms can be realized, e.g., by using the experimental
technique based on the electron microscopy \cite{exper_Ott_1,exper_Ott_2}. A narrow
electron beam is directed to one of the minima of the potential ionizing the atoms
and the latter are removed from the condensate. Alternatively one can use  a narrow
laser beam. In both cases, the interaction with the beam serves as a continuous
measurement tool and can be described in the framework of the standard Markovian
approximation~\cite{BP}. Introducing the probability $p\equiv p(k_1,\Delta t)$,
where $\Delta t$ is the  time interval and $k_m$ is a population of the $m$th well,
the single-atom removal event can be cast  as a quantum channel:
\begin{equation}
|k_1,k_2\rangle|0\ran_R\to\sqrt{p}|k_1-1,k_2\rangle|1\ran_R+\sqrt{1-p
}|k_1,k_2\rangle|0\ran_R,
\label{QC}\end{equation}
where the atoms are removed from well 1,
$|k_1,k_2\ran=\frac{(a_1^\dag)^{k_1}(a_2^\dag)^{k_2}}{\sqrt{k_1!k_2!}}|0\rangle$ is
the ket vector of the BEC state and $|j\ran_R$ describes the atom counter.
Introducing the atom removal rate $\Gamma$, we approximate $p(k_1,\Delta t)\approx
\Gamma k_1 \Delta t$ for small $\Delta t$, much less than the tunneling time
$t_\mathrm{QT} = {\hbar}/{J}$. In particular, in the experiments with the electron
microscopy, the atom removal rate is computed to be $\Gamma \approx
I\sigma_{tot}/e$~\cite{exper_Ott_2}, where $I$ is the current of the electron beam,
$e$ is the electric charge of the electron, and $\sigma_{ion}$ is the total
ionization cross section.

In terms of the reduced density matrix $\rho$, describing the condensate alone, the
quantum channel (\ref{QC}) is given by the Kraus superoperator representation $\rho
\to M_0\rho{M}^\dag_0 + M_1\rho{M}^\dag_1$, where for a small $\Delta t$ we have
$M_0\approx 1-\Gamma n_1\Delta t/2$ and $M_1\approx \sqrt{\Gamma \Delta t\,}a_1$.
This leads to the master equation in the Lindblad form
\eqb
\frac{d{\rho}}{d t} = -\frac{i}{\hbar} [H,\rho]
+\Gamma\left\{a_1\rho{a_1^\dag}-\frac{n_1}{2}\rho-\rho\frac{n_1}{2}\right\}.
\label{EQ5}
\eqe
The Lindblad operator $\mathcal{D}(\cdot)=a(\cdot)a^\dag_1-\{n_1,(\cdot)\}/2$ has
the eigenvalues $\lambda\in \{-N,-N+1/2,...,-1/2,0\}$, where $N$ is the total
number of atoms. The eigenstates are in the product form $\rho =
\rho^{(1)}\otimes\rho^{(2)}$, with $\rho^{(j)}$ corresponding to  the $j$th well.
Respectively, the right-hand side of Eq. (\ref{EQ5}), considered as a
superoperator, has eigenvalues in the form $\lambda = i\Delta E - \mu$, where
$\Delta E$ has the range of values  of the difference between  the energy levels of
the Hamiltonian (\ref{EQ1}), while $0\le \mu\le \Gamma N$. The only  stationary
state (i.e. attractor) is the zero eigenvalue eigenstate, which has $\rho^{(1)}_0 =
|0\rangle\langle0|$, that is, no atoms in the left well. The dissipation part is
also responsible, besides the removal of atoms, for the gradual loss of coherence
between the wells of the double-well trap, counteracting the effect of the quantum
tunneling. In its turn, the tunneling depends on the interactions between the atoms
\cite{MFDW,MFDW2,QHAM}, thus leading to interesting dissipation regimes governed by
the master equation (\ref{EQ5}).

\section{Noninteracting-atoms case}
\label{sec:noninteract}

Consider first the case of $U=0$, which can be achieved by making $a_s$ negligible
using the Feshbach resonance. Then Eq.~(\ref{EQ5}) can be solved explicitly by
using the adjoint equation (see, Ref.~\cite{BP})
\eqb
\frac{d\hat{A}}{d t} = \frac{i}{\hbar}[H,\hat{A}] +
\Gamma\left\{{a_1^\dag}\hat{A}{a}_1-\left(\frac{n_1}{2}\hat{A}+\hat{A}\frac{n_1}{2}\right)\right\}
\label{EQ6}
\eqe
for an  observable $\hat{A}$ [we distinguish  an operator solution of
Eq.~(\ref{EQ6}) with a hat]. First of all, for $\hat{a}_j(t)$ ($j=1,2$)
Eq.~(\ref{EQ6}) is solved by the ansatz $\hat{a}_j(t) = C_{j1}(t) a_1 +C_{j2}(t)
a_2$ with the initial condition $\hat{a}_j(0)=a_j$. This gives
\begin{eqnarray}
 && C_{11} =
\frac{\lambda_+  e^{\lambda_+t}- \lambda_- e^{\lambda_-t}}{2i \Omega } ,\quad
C_{22} =\frac{\lambda_+e^{\lambda_-t}-\lambda_-e^{\lambda_+t}}{2i \Omega }
\nonumber\\
&& C_{12} =C_{21}= \frac{J\left(e^{\lambda_+t}-e^{\lambda_-t}\right)}{2\hbar \Omega }
\nonumber
\label{EQ7}
\end{eqnarray}
where $\lambda_\pm = -\frac{\Gamma}{4}-\frac{i\vare}{2\hbar}\pm i \Omega$ and
$\Omega=
\sqrt{\frac{J^2}{\hbar^2}-\left[\frac{\Gamma}{4}+\frac{i\vare}{2\hbar}\right]^2}$.
The time dependence of the operators $\hat{a}_j^\dag(t)$ is given by the Hermitian
conjugated expressions. Moreover, the time dependence of an  arbitrary operator can
be found in terms of the operators $\hat{a}_j(t)$ and
${\hat{a}_{j^\prime}}^\dag(t)$. This is due to  the conditional ``Leibnitz rule''
for the dissipation part of Eq.~(\ref{EQ6}):
$\mathcal{D}^*[\hat{A}\hat{B}]=\mathcal{D}^*[\hat{A}]\hat{B}+\hat{A}\mathcal{D}^*[\hat{B}]$
valid when either $[a_1^\dag,\hat{A}]=0$ or $[a_1,\hat{B}]=0$. Hence, the
coefficients $C_{ij}(t)$ are sufficient to determine evolution of any observable.

Setting, for simplicity, $\vare=0$, and assuming that initially the condensate is
in the ground state $|\psi\rangle =
\frac{(a^\dag_1+a^\dag_2)^{N_0}}{\sqrt{2^{N_0}N_0!}}|0\rangle$ we obtain
\begin{eqnarray}
&&\langle N\rangle =e^{-\frac{\Gamma}{2}t}
\left[\frac{J^2}{(\hbar\Omega)^2}-\frac{\Gamma^2}{16\Omega^2}\cos(2\Omega
t)\right]N_0,
\label{EQ8}\\
&&\langle n_1-n_2\rangle = -e^{-\frac{\Gamma}{2}t} \frac{\Gamma}{4\Omega}
\sin(2\Omega t)N_0.
\label{n}\end{eqnarray}
When $\Gamma<4J/\hbar$, Eqs. (\ref{EQ8}) and (\ref{n}) describe the decaying  Rabi
oscillations with the decay rate $\Gamma/2$. Surprisingly, for $\Gamma\gg
4J/\hbar$, the dynamics is characterized by two different loss rates:  the initial
stage  with the rate $\Gamma/2$ and,  for times exceeding $1/\Gamma$, a
dramatically reduced dissipation rate   $\Gamma_\mathrm{QT} \approx
\frac{4J^2}{\hbar^2\Gamma}$. This can be explained as follows. Consider, for
simplicity, the case $\Gamma \gg J/\hbar$, that is,  when $\Gamma_{QT}\ll\Gamma$.
We get for $t\gg 1/\Gamma$:
\eqb
\langle n_{1}\rangle
\approx\frac{\Gamma_{QT}}{\Gamma}\frac{N_0}{2}e^{-\Gamma_{QT}t},\quad \langle
n_{2}\rangle \approx \frac{N_0}{2}e^{-\Gamma_{QT}t},
\label{EQn1n2}
\eqe
Observe that initially the two potential minima are equally populated ($\langle
n_j\rangle=N_0/2$). Equation  (\ref{EQn1n2}) shows that there few   atoms
($\Gamma_{QT}/\Gamma\ll1$)  in  well 1 after  the time scale $t\sim 1/\Gamma$, that
is, the system state is close to the zero-eigenvalue eigenstate  of the dissipation
part of Eq. (\ref{EQ5}). Further elimination of the atoms in this regime occurs via
the quantum tunneling from well 2, thus giving origin to the dramatically reduced
loss rate $\Gamma_\mathrm{QT}$.

The prevention of losses of atoms by a strong dissipation resembles the Zeno effect
observed experimentally in a different setup \cite{CPZeno}. Specifically,
expressing our $\Gamma_{QT}$ through the  tunneling frequency $\omega_R = 2J/\hbar$
we recover the decay rate $\Gamma_{QT} = \omega_R^2/\Gamma$ which appears in the
continuous Zeno effect of Ref. \cite{CPZeno}.

\section{The case of interacting atoms}
\label{sec:nonlin_Zeno}

In the nonlinear case one cannot solve Eq.~(\ref{EQ6}) exactly. The strategy now is
to use the mean-field approximation, valid in the limit  of a large number of
atoms. To this end we consider  the equations for the averaged quantities:
\begin{subequations}
\begin{eqnarray}
\label{EQ14_1}
& & \frac{d\lan n_1\ran}{dt}=i J\left(\lan a_1^\dag a_2\ran-\lan a_2^\dag
a_1\ran\right)-\Gamma\lan n_1 \ran
\\
\label{EQ14_3}
& &\frac{d\lan n_2\ran}{dt}=-i J\left(\lan a_1^\dag a_2\ran-\lan a_2^\dag
a_1\ran\right)
\\
\label{EQ14_4}
& &\frac{d }{dt}\left(\lan a_1^\dag a_2\ran-\lan a_2^\dag a_1\ran\right)
=2iJ\left(\lan n_1 \ran-\lan n_2 \ran\right) \nonumber
\\
& & +2iU\left(\lan n_1 a_1^\dag a_2\ran+\lan n_1 a_2^\dag a_1\ran-\lan
n_2a_1^\dag a_2\ran -\lan n_2 a_2^\dag a_1\ran\right)\nonumber\\
& &-\Gamma\left(\lan a_1^\dag a_2\ran-\lan a_2^\dag a_1\ran\right),
\end{eqnarray}
\end{subequations}
To obtain a closed system from Eqs. (\ref{EQ14_1})-(\ref{EQ14_4}) in the nonlinear
case one can decouple the fourth-order correlators as follows: $\langle
n_ja^\dag_{j^\prime}a_{j^{\prime\prime}}\rangle\approx \langle n_j\rangle\langle
a^\dag_{j^\prime}a_{j^{\prime\prime}} \rangle$. This procedure corresponds to the
mean-field approximation, that is, $N\to \infty$, widely used for description of
BEC and cold atoms  for a large number of particles (we have checked its validity
using the direct quantum Monte-Carlo simulations). The mean-field variables, $z$,
$\phi$ and $q$, correspond to the averaged quantities:
\eqb
z=\frac{ \langle n_1\rangle - \langle n_2\rangle }{  \langle n_1\rangle + \langle
n_2\rangle }, \quad  e^{i\phi}=\frac{\lan a_1^\dag a_2\ran}{\sqrt{ \langle
n_1\rangle\langle n_2\rangle }}, \; q = \frac{\langle n_1\rangle + \langle
n_2\rangle}{N_0}.
\eqe
They  satisfy the system
\begin{subequations}
\label{z_n}
\bee
&&\frac{dz}{d\tau}=-2\sqrt{1-z^2}\sin\phi-\frac{\gamma}{2}\left(1-z^2\right),
\label{EQ14A}
\\
&&\frac{d\phi}{d\tau}=2\frac{z}{\sqrt{1-z^2}}\cos\phi +\varepsilon+2\Lambda q z,
\label{EQ15}
\\
&&\frac{dq}{d\tau}=-\frac{\gamma}{2}q(1+z).
\label{EQ16}
\ene
\end{subequations}
Here we  use  the dimensionless time $\tau = t/t_{QT}$ and the normalized atom
removal rate $\gamma = \Gamma t_{QT} = \hbar\Gamma/J$.    The parameter $\Lambda =
UN/J$  characterizes  the atomic interactions in the condensate \cite{MFDW,MFDW2}.

\begin{figure}[htb]
\begin{center}
\epsfig{file=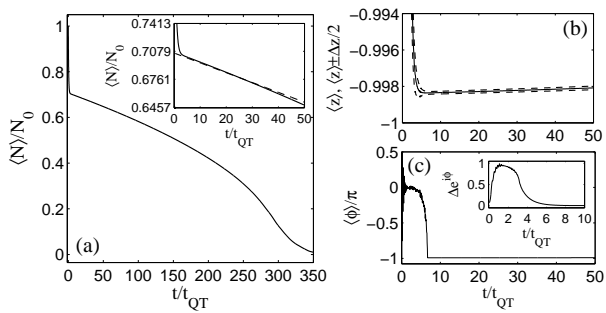,width=0.45\textwidth} \caption{The dissipation dynamics of
the self-trapping state. Here $\Lambda = 25$, $\vare=0$ and $\gamma =2$. (a) The
ratio of the number of the remaining atoms. The inset compares the analytical
estimate for the nonlinear dissipation rate (dashed line) with the numerical
result.  (b) The normalized population imbalance $\langle z \rangle$   (the dashed
lines show the dispersion). (c) The evolution of the phase $\langle\phi \rangle$;
the inset gives  the phase fluctuations $\Delta e^{i\phi}$. The initial state is a
Gaussian distribution in the phase space with $\langle z\rangle=-0.5$ and
$\langle\phi\rangle=0$, and dispersions $\Delta z = 0.05$ and $\Delta e^{i\phi} =
0.1$. }
\label{FG2}
\end{center}
\end{figure}

We have checked, by comparing with the direct quantum Monte-Carlo simulations, that
solutions of Eqs.~(\ref{EQ14A})-(\ref{EQ16}) averaged  in the classical phase space
give an excellent agreement with Eq.~(\ref{EQ5}) for $N\sim100$ (whereas a good
agreement is observed already for $N\sim 10$). The essentially linear dynamics is
observed for the interaction strength $\Lambda\lesssim 1$. However, for strong
interactions ($\Lambda\gg1$)  the  self-trapping  state \cite{MFDW,MFDW2}  features
a dramatic reduction of the atom loss rate with increasing interaction strength
(see Fig.~\ref{FG2}; the self-trapping state is in well 2). The effect does not
depend on the interaction type (attractive, $\Lambda<0$, or repulsive, $\Lambda>0$)
and appears for any value of $\Gamma$.

This effect can be understood as follows. First, one has to distinguish between the
two types of the self-trapping states in the Hamiltonian system: the running-phase
states with $\phi(t)\propto t$ and the fixed-phase states (see Refs.
\cite{GO,MFDW2,STpaper}). There are the following fixed-phase self-trapping states:
$\cos\phi = -\mathrm{sgn}(\Lambda)$ and $z = \pm\sqrt{1-\Lambda ^{-2}}$. In the
dissipative case, $z=-1$ defines an invariant subset of system (\ref{z_n})
representing the mean-field description of the zero-eigenvalue eigenstate of the
Lindblad term in Eq. (\ref{EQ5}), corresponding to the location of all atoms in
well 2. Accordingly, the fixed-phase self-trapping state with $z<0$ of the
Hamiltonian system shows  the new nonlinear dissipation rate, right after the
self-trapping is significantly enhanced by the dissipation: $z$ decreases to $-1$,
while the phase becomes equal to an odd ($\Lambda>0$) or even ($\Lambda<0$)
multiple of $\pi$ (the phase fluctuations, described by the expression $\Delta
e^{i\phi}=[1-|\langle e^{i\phi}\rangle|^2]^{1/2}$ \cite{STpaper}, decay to zero).

Further insight can be gained by visualizing the atomic interactions as an
effective common potential experienced by the condensed atoms. Indeed, in the
self-trapping state $\langle n_1\rangle \ll \langle n_2\rangle$; hence, the
nonlinear term in the averaged Hamiltonian (\ref{EQ1}) can be simplified
\begin{eqnarray}
\label{expan_nonlin}
U\left[\langle n_1^2\rangle + \langle n_2^2\rangle \right]\approx U\langle
n_2\rangle^2 \approx U\langle N\rangle  - 2U \langle N \rangle \langle n_1\rangle,
\end{eqnarray}
where we have used that the fluctuations of  $z$ are small (see Fig. \ref{FG2},
that is,  $\langle n^2_2\rangle\approx \langle n_2\rangle^2$. The  total number of
atoms evolves adiabatically, since the dissipation rate is very small. Then, the
last term in Eq. (\ref{expan_nonlin}), which is proportional to $\lan n_1\ran$, is
but a simple renormalization of $\vare$: $\vare\to\vare - 2U\langle N\rangle$ in
Eq. (\ref{EQ5}) [respectively, (\ref{EQ15})]. This allows one to derive the
nonlinear dissipation rate of the self-trapping state. Considering the symmetric
trap ($\vare=0$) and setting $\vare_\mathrm{NL}\equiv -2U\langle N\rangle $,
provided that $\hbar^2\Gamma^2 +4\vare^2_\mathrm{NL}\gg J^2$, we get for  $t
\gtrsim 1/\Gamma$
\eqb
\Gamma_{\mathrm{NL}}\approx \frac{4J^2\Gamma}{\hbar^2\Gamma^2 +
4\vare_\mathrm{NL}^2}.
\label{NLGAM}\eqe
This estimate turns out to be in excellent agreement with the numerical results (at
the initial stage of the decay of the self-trapping state), as it is shown in
Fig.~\ref{FG2} (if one uses in Eq. (\ref{NLGAM}) the actual numerical number of
atoms $\langle N\rangle$ remaining in the self-trapping state). In the regime with
$\vare_\mathrm{NL}\gg\hbar\Gamma$ the nonlinear dissipation rate (\ref{NLGAM})
reduces to $\Gamma_{\mathrm{NL}}\approx {J^2\Gamma}/{\vare_\mathrm{NL}^2}$. The
preceding  inequality condition also means that $\Gamma_{\mathrm{NL}}\ll \Gamma$,
that is, the actual dissipation rate is dramatically reduced. This inhibition of
the losses for $\Lambda \gg1$ can be viewed as  a nonlinear Zeno effect which, in
contrast to the usual Zeno effect \cite{Zeno}, appears for arbitrary $\Gamma$.

To conclude this section let us make two remarks on the nonlinear decay rate given
by  Eq. (\ref{NLGAM}). First, we notice that while the analogy between the effect
of two-body interactions and the bias $\varepsilon$ in pure linear system worked
well in the estimate for the decay rate in of the nonlinear Zeno effect, we
emphasize that the role of these factors in the dynamics under consideration is
very different. The linear bias is an external factor which defines the constant
decay rate of the linear Zeno effect, that is,  of a condensate of noninteracting
atoms. The ``induced bias" due to the two-body interactions, that is,  $-2U\langle
N\rangle$, by itself is a dynamical quantity: It slowly changes with time,
resulting, after all, in the change of the decay rate. In other words, after a
sufficiently long time, sufficient for a significant loss of atoms, the nonlinear
decay rate  will coincide with the  linear one. Second, the decay rate given in Eq.
(\ref{NLGAM}) has a broader application than just giving the decay of the
self-trapping  state (thus, it is a \textit{new} nonlinear effect, different from
the self-trapping itself). For instance, it gives a correction due to the
nonlinearity to the  decay rate of the linear   Zeno effect also for a small
$\Lambda$. Indeed, there are two conditions of validity of Eq. (\ref{NLGAM}):
$\hbar^2\Gamma^2 +4\vare^2_\mathrm{NL}\gg J^2$, satisfied also by taking large
$\Gamma\gg J/\hbar$ (and an arbitrary $\Lambda$), and that there few atoms in well
1, which is the stage of the  Zeno effect.

\section{Quantum switching induced by dissipation}
\label{sec:switch}

\begin{figure}[floatfix]
\begin{center}
\epsfig{file=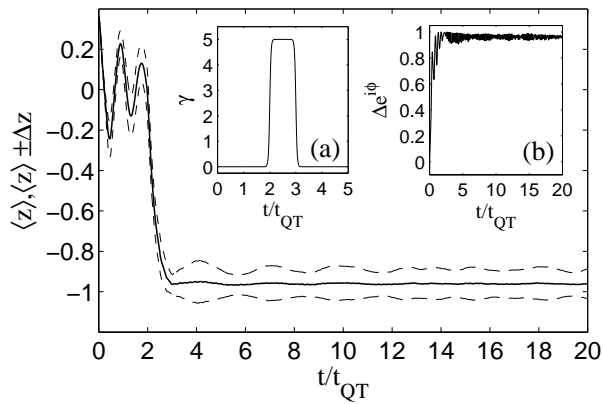,width=0.45\textwidth} \caption{ Dissipation-induced
self-trapping state. Here $\Lambda = 25$ and an initial Gaussian distribution in
the phase space with $\langle z\rangle =0.35$, $\langle\phi\rangle=0$, $\Delta z =
0.03$ and $\Delta e^{i\phi} = 0.1$ was used. In the main panel, the solid line
gives the average and the dashed lines give the dispersion,  $\gamma(t)$ is given
in inset (a) and in inset (b) the phase fluctuations are shown.  }
\label{FG3}
\end{center}
\end{figure}

Quantum switching  can be induced by simply removing atoms for a short time in a
switch-on manner. For instance, application of the atom removal for a time interval
on the order of $1/\Gamma$ draws the system close to   the self-trapping state with
$z\approx-1$. This is illustrated in Fig. \ref{FG3}, where about 50\% of initially
loaded atoms remain in a self-trapping state, whereas until the action of
dissipation, the system was in the Josephson oscillations regime.

\begin{figure}[floatfix]
\begin{center}
\epsfig{file=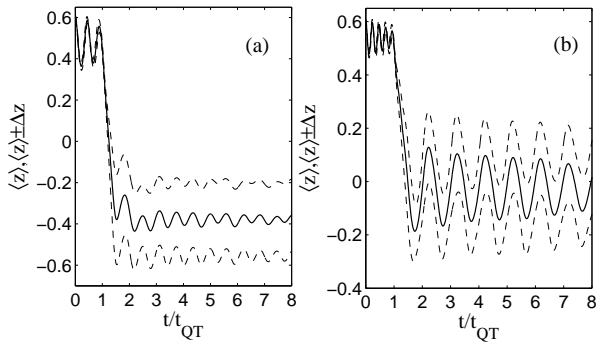,width=0.45\textwidth} \caption{Dissipation-induced
switching. In (a),  $\Lambda = 15$ and the dissipation acted in the interval
$\Delta t= 0.5t_\mathrm{QT}$ leaving 50\% of atoms in the system. In (b), $\Lambda
= 25$ and the dissipation is switched for an interval $\Delta t=0.3t_\mathrm{QT}$
leaving 55\% of atoms in the system. The dissipation is switched on at $t/t_{QT}=1$
and has $\gamma_\mathrm{max} =3$. The solid and dashed lines denote the average and
dispersion, respectively.  The initial Gaussian distribution has $\langle
z\rangle=0.6$, $\langle\phi\rangle=0$, $\Delta z = 0.02$ and $\Delta e^{i\phi} =
0.2$. }
\label{FG4}
\end{center}
\end{figure}

By directing the dissipation tool to the well where the self-trapping state is
located and by varying the time of application of the dissipation and the other
parameters, for example,   $\Lambda$, one can obtain the switching between the
self-trapping states in the two wells of the double-well trap, Fig. \ref{FG4}(a),
or induce the switching to the macroscopic quantum tunneling regime, Fig.
\ref{FG4}(b).  The nonadiabatic variation of the dissipation  induces large phase
fluctuations, $\Delta e^{i\phi}=[1-|\langle e^{i\phi}\rangle|^2]^{1/2}\to 1$, see
Fig. \ref{FG3}(b),  in contrast to the continuous action of a constant dissipation,
where the phase fluctuations rapidly decay.

\section{Conclusion}

Selective  removal of atoms by a directed external beam of particles  can be viewed
as a continuous measurement or, alternatively, as an induced controlled dissipation
of a mesoscopic system. Combining these, apparently different interpretations of
the interaction between a given mesoscopic system and an external system considered
as a reservoir, with the nonlinearity due to interatomic interactions, opens
remarkable possibilities for observation of the fundamental quantum phenomena in
open systems on   one hand, and on  the other hand, for preparation  and controlled
manipulation of mesoscopic systems, for example, a Bose-Einstein condensate. More
specifically, we have shown that while the controlled dissipation attenuates losses
of atoms (already known phenomenon, interpreted also as the celebrated Zeno
effect), when combined with the strong atomic interactions it results in an
essentially new regime, where the atomic decay rate is practically zero,  due to
emergence of a quasi-self-trapping state. Such a state still loses atoms  (unlike
the case of the standard self-trapping state at a fixed number of atoms) and in
this sense, the nonlinear Zeno effect can be viewed as a signature of the
self-trapping state. This new effect of drastic attenuation of the losses of atoms
by the atomic interactions, which appears for any  value of the dissipation
parameter, can be viewed as the nonlinear Zeno effect. Moreover, application of the
dissipation during a short interval  of time opens new possibilites for control
over the condensate, for example,  inducing the switching   between two
self-trapping states in the two wells of the double-well trap or between a
self-trapping state and the macroscopic quantum tunneling regime.

We have considered only the simplest nontrivial trap, the double-well potential. In
general,  one should expect much  richer  behavior  in the multiwell potentials
and, moreover, in the case of  BEC loaded in the optical lattices, manipulated by a
local  dissipation.

The macroscopic nature of the described phenomena  suggests that they can be
observed in any nonlinear physical system which is described by the nonlinear
Schr\"odinger-like equation and has at least two different equilibria, one of which
is subjected to dissipative losses, that is, in a fairly generic setup. For
instance, there is  a very close analogy with the nonlinear optics of Kerr media,
in particular, the nonlinear optical fibers (say the standard Kerr fibers or
hollow-core fibers filled with atomic gases), suggesting that the nonlinear Zeno
effect can be observed also in the realm of nonlinear optics in the currently
available experimental settings.

\section{Acknowledgements}
The authors acknowledge the financial support of the CNPq and FAPESP of Brazil.

\end{document}